\documentstyle[12pt]{article}
\newlength{\extraspace}
\newlength{\extraspaces}
\renewcommand{\theequation}{\thesection.\arabic{equation}}
\newcommand{\lae}{\begin{array}{c}\,\sim\vspace{-21pt}\\< \end{array}}

\newcommand{\be}{\begin{equation}}
\newcommand{\ee}{\end{equation}}
\newcommand{\bear}{\begin{eqnarray}}
\newcommand{\eear}{\end{eqnarray}}
\newcommand{\ba}{\begin{array}}
\newcommand{\ea}{\end{array}}
\newcommand{\mlq}{\mbox{${M_\inl^q}^2$}}
\newcommand{\mru}{\mbox{${M_\inr^u}^2$}}
\newcommand{\mrd}{\mbox{${M_\inr^d}^2$}}
\newcommand{\mluc}{\mbox{$({M_\inl^q}^2)_{12}$}}
\newcommand{\mlut}{\mbox{$({M_\inl^q}^2)_{13}$}}
\newcommand{\mlct}{\mbox{$({M_\inl^q}^2)_{23}$}}
\newcommand{\mlcts}{\mbox{$({M_\inl^q}^2)_{23}^2$}}
\newcommand{\mluu}{\mbox{$({M_\inl^q}^2)_{11}$}}
\newcommand{\mlcc}{\mbox{$({M_\inl^q}^2)_{22}$}}
\newcommand{\mltt}{\mbox{$({M_\inl^q}^2)_{33}$}}

\newcommand{\mrut}{\mbox{$({M_\inr^u}^2)_{13}$}}
\newcommand{\mrct}{\mbox{$({M_\inr^u}^2)_{23}$}}

\newcommand{\mrdb}{\mbox{$({M_\inr^d}^2)_{13}$}}
\newcommand{\mrsb}{\mbox{$({M_\inr^d}^2)_{23}$}}
\newcommand{\mrdd}{\mbox{$({M_\inr^d}^2)_{11}$}}
\newcommand{\mrss}{\mbox{$({M_\inr^d}^2)_{22}$}}
\newcommand{\mlrb}{\mbox{$({M_{\inl\inr}^d}^{\!\!\! 2})_{33}$}}
\newcommand{\mlrij}{\mbox{$({M_{\inl\inr}^d}^{\!\!\! 2})_{ij}$}}
\newcommand{\mlij}{\mbox{$({M_\inl^q}^2)_{ij}$}}
\newcommand{\mruij}{\mbox{$({M_\inr^u}^2)_{ij}$}}
\newcommand{\mrdij}{\mbox{$({M_\inr^d}^2)_{ij}$}}
\newcommand{\mlucp}{\mbox{$({M_\inl^u}^2)_{12}^\prime$}}

\newcommand{\mldsp}{\mbox{$({M_\inl^d}^2)_{12}^\prime$}}
\newcommand{\mldbp}{\mbox{$({M_\inl^d}^2)_{13}^\prime$}}
\newcommand{\mlsbp}{\mbox{$({M_\inl^d}^2)_{23}^\prime$}}
\newcommand{\mrucp}{\mbox{$({M_\inr^u}^2)_{12}^\prime$}}

\newcommand{\mrdsp}{\mbox{$({M_\inr^d}^2)_{12}^\prime$}}
\newcommand{\mrdbp}{\mbox{$({M_\inr^d}^2)_{13}^\prime$}}
\newcommand{\mrsbp}{\mbox{$({M_\inr^d}^2)_{23}^\prime$}}
\newcommand{\mllij}{\mbox{$({M_\inl^l}^2)_{ij}$}}
\newcommand{\mreij}{\mbox{$({M_\inr^e}^2)_{ij}$}}
\newcommand{\mU}{\mbox{$m^{u}$}}

\newcommand{\muu}{\mbox{$m^{u}_{11}$}}
\newcommand{\muc}{\mbox{$m^{u}_{12}$}}
\newcommand{\mcu}{\mbox{$m^{u}_{21}$}}
\newcommand{\mcc}{\mbox{$m^{u}_{22}$}}
\newcommand{\mtt}{\mbox{$m^{u}_{33}$}}
\newcommand{\muij}{\mbox{$m^{u}_{ij}$}}
\newcommand{\fuij}{\mbox{$f^{u}_{ij}$}}
\newcommand{\fucc}{\mbox{$f^{u}_{22}$}}
\newcommand{\eql}{\mbox{$\epsilon^{q}_{\inl}$}}
\newcommand{\eur}{\mbox{$\epsilon^{u}_{\inr}$}}
\newcommand{\edr}{\mbox{$\epsilon^{d}_{\inr}$}}
\newcommand{\vul}{\mbox{$V^{u}_{\inl}$}}
\newcommand{\vur}{\mbox{$V^{u}_{\inr}$}}
\newcommand{\vdl}{\mbox{$V^{d}_{\inl}$}}
\newcommand{\vdr}{\mbox{$V^{d}_{\inr}$}}
\newcommand{\mds}{\mbox{$m^{d}_{12}$}}
\newcommand{\mSs}{\mbox{$m^{d}_{22}$}}
\newcommand{\msb}{\mbox{$m^{d}_{23}$}}
\newcommand{\mbb}{\mbox{$m^{d}_{33}$}}
\newcommand{\mct}{\mbox{$m^{u}_{23}$}}
\newcommand{\up}{\mbox{$p$}}
\newcommand{\down}{\mbox{$m$}}
\newcommand{\bet}{\mbox{$\beta^u$}}
\newcommand{\De}{\mbox{$\Delta$}}

\newcommand{\suu}{\mbox{$SU(2)\times U(1)$}}

\newcommand{\ew}{\mbox{$SU(2)_{W}\times U(1)_{Y} \ $}}
\newcommand{\hc}{{\rm h.c.}}
\newcommand{\gev}{\mbox{\rm GeV}}
\newcommand{\tev}{\mbox{\rm TeV}}

\newcommand{\trei}{{\scriptscriptstyle 3}}
\newcommand{\unu}{{\scriptscriptstyle 1}}
\newcommand{\doi}{{\scriptscriptstyle 2}}
\newcommand{\suw}{\mbox{$SU(2)_W \ $}}

\newcommand{\tcsm}{\mbox{$SU(N_{TC})\times SU(3)_{C}\times
SU(2)_{W}\times U(1)_{Y} \ $}}
\newcommand{\tcsma}{\mbox{$SU(N_{TC})\times $}}
\newcommand{\tcsmb}{\mbox{$SU(3)_{C}\times $}}
\newcommand{\tcsmc}{\mbox{$SU(2)_{W}\times $}}
\newcommand{\tcsmd}{\mbox{$U(1)_{Y} $}}
\newcommand{\inl}{{\scriptscriptstyle L}}
\newcommand{\inr}{{\scriptscriptstyle R}}
\newcommand{\inlr}{{\scriptscriptstyle LR}}
\newcommand{\fir}{\mbox{$\tilde{\phi}_{\inr}$}}
\newcommand{\fil}{\mbox{$\tilde{\phi}_{\inl}$}}
\newcommand{\SM}{{\scriptscriptstyle SM}}
\newcommand{\mfit}{\mbox{$ \cM_{\phi}$}}
\newcommand{\mfis}{\mbox{$ \cM_{\phi}^2$}}
\newcommand{\mfil}{\mbox{$ (M^2_{\tilde{\phi}})_{\inl\inl}$}}
\newcommand{\mfir}{\mbox{$ (M^2_{\tilde{\phi}})_{\inr\inr}$}}
\newcommand{\mfilr}{\mbox{$ (M^2_{\tilde{\phi}})_{\inlr}$}}
\newcommand{\mfilrs}{\mbox{$ (M^2_{\tilde{\phi}})^2_{\inlr}$}}
\newcommand{\chil}{\mbox{$\tilde{\chi}_{\inl}$}}
\newcommand{\chir}{\mbox{$\tilde{\chi}_{\inr}$}}
\newcommand{\mchit}{\mbox{$ \cM_{\chi}$}}
\newcommand{\mchis}{\mbox{$ \cM_{\chi}^2$}}
\newcommand{\mchilr}{\mbox{$ (M^2_{\tilde{\chi}})_{\inlr}$}}

\newcommand{\Mkl}{\mbox{$ {M_k^q}_{\! \inl}^2$}}
\newcommand{\Mil}{\mbox{$ {M_i^q}_{\!\! \inl}^2$}}

\newcommand{\Mlr}{\mbox{$ {M_l^u}_{\!\!\! \inr}^2$}}
\newcommand{\Mir}{\mbox{$ {M_i^u}_{\!\!\! \inr}^2$}}

\newcommand{\ms}{\mbox{$ \cM^2_s$}}
\newcommand{\mss}{\mbox{$ \cM_s$}}
\newcommand{\dg}{\mbox{$ \delta_{\tilde{g}} $}}
\newcommand{\flavor}{\mbox{$ SU(3)\times U(1)$}}

\newcommand{\fla}{\mbox{$ [SU(2)\times U(1)]^5$}\ }
\newcommand{\cM}{{\cal M}} \newcommand{\cO}{{\cal O}}
\begin{document}
\pagestyle{empty}
\begin{titlepage}
\begin{flushright}
{BUHEP-95-12 \\ hep-ph/9504399 \\
April 26, 1995 \\ Revised June 2, 1995}
\end{flushright}
\vspace{24pt}
\begin{center}
{\LARGE Fermion Masses without Higgs:\\ \vspace{5pt} A Supersymmetric
Technicolor Model}\\
\vspace{40pt}
{\large
{\rm Bogdan A. Dobrescu}}\footnote{e-mail address:
dobrescu@budoe.bu.edu}
\vspace*{0.5cm}

{\large
{\it Department of Physics, Boston University \\
590 Commonwealth Avenue, Boston, MA 02215, USA}
}
\vskip 2.1cm
\rm
\vspace{25pt}
\vspace{12pt}
\end{center}
\baselineskip=18pt

\begin{abstract}

{\normalsize
We propose a supersymmetric technicolor model in which the electroweak
symmetry breaking is communicated to the quarks and leptons by
technicolored $SU(2)_W$-singlet scalars. When the technifermions
condense, the quarks and leptons of the third generation acquire mass.
The fermions of the other generations do not couple to the technicolored
scalars but they receive masses from radiative corrections involving
superpartners. As a result, the mass hierarchy between the fermion
generations arises naturally. The model predicts the CP asymmetries in
$B$ meson decays and in $\Delta S = 1$ transitions to be smaller by two
orders of magnitude than the ones predicted in the Standard Model.
}

\end{abstract}

\vfill
\end{titlepage}

\baselineskip=18pt
\pagestyle{plain}
\setcounter{page}{1}

\section{Introduction}
\label{sec:intro}
\setcounter{equation}{0}

The Higgs doublet has a double role in the Standard Model (SM):
to break the electroweak symmetry spontaneously and to give mass
to the fermions. The latter offers no
explanation for the pattern of masses of the quarks and leptons.
The mixing angles of the quarks and the CP violation phase are
also free parameters in the SM. Moreover, the arbitrariness in
the phase of the quark mass determinant is one of the sources of the
strong CP problem \cite{cheng}.

The existence of the light Higgs boson in the SM
is unnatural \cite{thooft} because of the quadratic divergences
in the scalar self-energy. This naturalness problem
can be solved while maintaining fundamental scalars in a theory
with supersymmetry (SUSY) broken softly \cite{kane}.
However, the simple structure of the Higgs sector of the SM
is lost in the Supersymmetric Standard Model (SSM) where there
is need for two Higgs doublets to provide mass for both up-type
and down-type quarks. Furthermore, in the SSM the constraints from
flavor-changing neutral currents (FCNC) require a high
degeneracy between the squarks with the same charge \cite{masiero},
which could occur only if strong assumptions are imposed
\cite{hall, flavor, leigh}.

In technicolor models \cite{farhi} the electroweak symmetry is broken
dynamically and there is no need for a Higgs doublet
provided a mechanism for fermion mass generation is found.
In extended technicolor (ETC) \cite{lane} the \ew symmetry
breaking is transmitted to the quarks and leptons by gauge bosons.
Since this is a renormalizable theory without fundamental scalars,
the naturalness problem is avoided.
However, the ETC models that give rise to correct fermion masses
have troubles with large FCNC, light pseudo-Goldstone bosons
and electroweak precision measurements.
Significant attempts to construct realistic ETC models were made
recently \cite{terning} but phenomenological problems remain to be
solved \cite{steve, balaji}.

Although technicolor was introduced as a mechanism for electroweak
symmetry breaking which does not depend on the existence of fundamental
scalar fields, it is possible to construct technicolor models
containing fundamental scalars.
Simmons \cite{simmons} considered a technicolor model with the
ordinary fermions receiving mass due to a massive scalar doublet
which couples to the technicolor condensate.
The phenomenology of this model is acceptable for a large range
of parameters \cite{simmons,carone}. Another possibility is that the
scalar doublet is massless while the physical scalar states acquire
mass from radiative corrections \cite{georgi}.
The naturalness problem does not appear in technicolor with a scalar
doublet provided the scalar is a composite state, such as a
fermion-antifermion state bound by fine-tuned ETC interactions
\cite{sekhar}. Samuel \cite{samuel} considered a
supersymmetric\footnote{Earlier attempts of combining SUSY and
technicolor can be found in Ref.~\cite{tcsusy}.}
version of this model, called bosonic technicolor,
which avoids most of the problems of technicolor and of the SSM
\cite{samuel,dine,kagan}. As in the case of the SM, these models with
scalar doublets offer no insight into the structure of the quark and
lepton mass matrices. An exception is a multi-Higgs model \cite{kagan}
with Yukawa couplings controlled by horizontal symmetries.

An explanation for the peculiar pattern of fermion masses might require
a mechanism for fermion mass generation not based on the Yukawa
couplings of the Higgs doublet. In the mechanism for generating
dynamical fermion masses proposed by Kaplan \cite{kaplan}
the exchange of technicolored $SU(2)_W$-singlet scalars induces
four-fermion effective interactions involving three technifermions and
one ordinary
fermion. As a result, the ordinary fermions contain an admixture of
technibaryon and acquire mass. A hierarchy of masses is produced but the
model predicts unacceptable FCNC and tree level contributions to the
$\rho$ parameter.

A different attempt to construct a realistic model, in which the
exchange of $SU(2)_W$-singlet techniscalars induces four-fermion
interactions between two ordinary fermions and two technifermions,
is due to Kagan \cite{kagan1}. In this model there are two doublets of
technifermions such that the techniscalar exchange contributions to the
fermion mass matrices have rank two. Therefore, only two generations
are massive at tree level. SUSY is necessary in this model in order to
avoid the naturalness problem but also it offers a source of radiative
masses for the fermions of the first generation.
The hierarchy between the second and third generation should be put in
by hand, as in the SM. However, since the fermion masses are quadratic
in Yukawa coupling constants, the fine-tuning of the Yukawa couplings
is less problematic in the model of Kagan than in the SM.
Phenomenological issues associated with $SU(2)_W$-singlet techniscalars
and different scenarios for quark mass generation
in non-SUSY theories are discussed in Ref.~\cite{kagan2}.

\vspace{6pt}
In this paper we propose a model in which the mass hierarchy between the
three generations of quarks and leptons arises naturally.

The model has several features in common with the model of Kagan.
There is no Higgs doublet and the \ew symmetry is broken by technicolor
interactions.
We introduce technicolored scalar fields which are $SU(2)_W$-singlets,
in order to couple the ordinary fermions to the technifermions.
When the  technifermions condense, the quarks and leptons acquire mass.
Since there are fundamental scalars in the model, their masses should
be protected by SUSY against quadratic divergences.

However, our model is more economic and more natural.
There is only one doublet of technifermions.
The flavor structure of this supersymmetric technicolor (SUSY-TC)
model leads to a realistic pattern of fermion masses. The reason is that
the Yukawa couplings of the techniscalars
can provide mass only for one generation of fermions while the other
two generations acquire smaller masses due to radiative corrections
involving gauginos, squarks and sleptons.
Such radiative fermion masses were discussed previously
\cite{lahanas,hall,kagan} but in those cases the ``chirality-flip''
mixing of squarks or sleptons was produced by Higgs couplings.
In our model the interaction of the squarks and sleptons with
the technifermions is at the origin of the chirality-flip mixing
(a similar mechanism is employed in Ref.~\cite{kagan1}).
The hierarchy between the second and first generations of fermions is
dictated by the structure of the squark and slepton mass matrices,
which, in turn, is suggested by the constraints on FCNC.
So far, the model has a viable phenomenology with distinctive
low energy predictions regarding the fermions of the third generation.

In Section 2 we describe the model.
We estimate the fermion masses in Section 3.
In Section 4 we discuss the constraints on squark masses from FCNC
and we study CP violation effects.
The main ideas are summarized in Section 5.

\section{The Model}
\label{sec:model}
\setcounter{equation}{0}

The gauge group is that of a minimal technicolor model:
\tcsma \tcsmb \tcsmc \\ \tcsmd.
The only source of electroweak symmetry breaking is the vacuum
expectation value of the technifermion bilinear. This condensate
couples to the weak gauge bosons which become massive. An ordinary
fermion has to couple to the condensate in order to acquire mass.
This can be done, as we will show in the discussion of quark and lepton
masses, by introducing an $SU(2)_W$-singlet scalar which has Yukawa
interactions with the ordinary fermion and a technifermion. Such
vertices are allowed by Lorentz invariance and gauge symmetry if
the left-handed technifermions are $SU(2)_W$-singlets and the
right-handed technifermions form doublets. To minimize the radiative
electroweak correction parameter $S$\ \cite{peskin} we introduce only
one doublet of technifermions.

We consider a low energy theory with global $N = 1$\ SUSY broken softly.
The Yukawa interactions of the technifermions
with the ordinary fermions appear in the superpotential which is
expressed only in terms of left-handed chiral superfields.
Thus, it is not possible to have the same scalar involved in the Yukawa
interactions of both left-handed and right-handed fermions.
However, the scalar superpartners of
the left-handed and right-handed components of an $SU(2)_W$-singlet
technifermion couple, respectively, to the left-handed and
the right-handed fermions; their mixing induces fermion masses.
The same pair of techniscalars couples to both the up-type and
down-type quarks. The gauge symmetry requires a different pair of
techniscalars to couple to the charged leptons.

The charges of the technicolored particles are uniquely determined by
imposing hypercharge conservation and the cancellation of the gauge
anomalies. The technicolored chiral superfields and their \tcsm
representations are listed below; the doublet of technifermions
responsible for electroweak symmetry breaking is contained in
\be
\Upsilon  \; : \; (N_{TC}, 1, 2)_0 \; , \hspace{.7cm}
\left\{ \ba{l} \up_{\inl} \; : \;  ({\bar{N}}_{TC}, 1, 1)_{-1} \\ [0.15cm]
\down_{\inl} \; : \;  ({\bar{N}}_{TC}, 1, 1)_1
\ea \right.~,
\label{e1}
\ee
where $\Upsilon = \left(\ba{l}\up^c\\ \down^c\ea\right) $\ ;
the scalar components of the $SU(2)_W$-singlet superfields
\bear
\chi_{\inl}  \; : \; ({\bar{N}}_{TC}, 1, 1)_1 \; & , & \;  \;
\chi^c  \; : \; (N_{TC}, 1, 1)_{-1}~, \nonumber \\ [0.15cm]
\phi_{\inl}  \; : \;
        ({\bar{N}}_{TC}, \bar{3}, 1)_{-\frac{1}{3}} \; & , & \; \;
\phi^c  \; : \; (N_{TC}, 3, 1)_{\frac{1}{3}}
\label{e2}
\eear
communicate the electroweak symmetry breaking to the leptons and quarks,
respectively. The superscript $c$\ denotes the charged conjugated
superfields. The techni-singlet chiral superfields are those of the MSSM
without the Higgs sector:
\bear
& & q_i = \left(\ba{l}u_{i_L} \\ d_{i_L}\ea\right) \; ,
\hspace{.7cm} u_i^c \: , \; d_i^c \: , \nonumber \\ [0.15cm]
& & l_i = \left(\ba{l}{\nu}_{i_L} \\ e_{i_L}\ea\right) \; ,
\hspace{.7cm} e_i^c~,
\label{e3}
\eear
where $i = 1,2,3$ is a generation index.

We mention that in a model with one family of technifermions
it is enough to introduce only one pair of techni-scalars which
are $SU(2)_W$-singlets in order to give masses to both quarks and
leptons. The drawback of such a model is that it contains four
doublets which produce a large contribution to the $S$\ parameter
\cite{peskin}. Note, also, that in the
one-doublet technicolor model presented here there are no
pseudo-Goldstone bosons. In more complicated models, such as the
two-doublet model of Kagan \cite{kagan1}, there are pseudo-Goldstone
bosons which require additional fields and interactions in order
to become massive.

The supersymmetric part of the Lagrangian contains kinetic terms for all
the fields, four-scalar interactions proportional to the gauge
coupling constants, Yukawa interactions of the fermion and scalar of
each chiral supermultiplet with the associated gauginos, and the
superpotential. In addition, there are soft SUSY breaking terms
\cite{grisaru} consisting of bilinear and trilinear scalar terms and
mass terms for the gauginos. Only gauge invariant terms which conserve
the baryon number $B$\ and the lepton number $L$\ are allowed in the
Lagrangian. We assign the following $B$\ and $L$\ numbers
to the technicolored chiral superfields:
\bear
\up_{\inl}, \, \down_{\inl} \: : \; L = 0 \: , \: B = 0
\nonumber \\
\chi_{\inl} \: : \; L = - 1 \: , \: B = 0
\nonumber \\
\phi_{\inl}  \: : \; L = 0 \: , \: B = - 1
\label{e4}
\eear

Apparently, the superpotential includes interactions of all three
generations of techni-singlet chiral superfields with the technicolored
superfields. However, after performing an appropriate
unitary transformation in the flavor space, only one generation (the
third one, by definition) couples to the
technicolored superfields; this is possible because the superpotential
is linear in the quark and lepton superfields.
Therefore, the most general superpotential is given by
\bear
\vspace{0.2cm}
W &=&
- \,
C_q\epsilon^{\alpha\beta}q_{\trei_{\alpha}}\Upsilon_{\beta}\phi_{\inl}
- C_t u_{\trei}^c \down_{\inl}\phi^c + C_b d_{\trei}^c\up_{\inl}\phi^c
- C_l\epsilon^{\alpha\beta}l_{\trei_{\alpha}}\Upsilon_{\beta}\chi_{\inl}
+ C_{\tau} e_{\trei}^c \up_{\inl} \chi^c
\nonumber \\ [0.15cm]
&&+ \, m_\chi\chi_{\inl}\chi^c +
m_\phi\phi_{\inl}\phi^c + \hc~,
\vspace{0.1cm}
\label{e5}
\eear
where $\alpha, \beta$\ are \suw indices, $\epsilon^{\alpha\beta}$\
is an antisymmetric tensor, $m_\chi$\ and $m_\phi$\ are mass parameters.
The signs in front of the trilinear terms correspond to positive
coupling constants $C_f (f = q,t,b,l,\tau)$\ in the expressions for
fermion masses.

Expressed in terms of scalar and fermion components,
the superpotential consist of Yukawa couplings,
four-scalar operators, and mass terms for the $SU(2)_W$-singlet
technicolored fermions and scalars.

The flavor redefinition we performed to obtain the superpotential
given by Eq.~(\ref{e5}) is an $[\flavor]^5$\ transformation where
there is an \flavor\ factor for each of the five chiral superfields
shown in Eq.~(\ref{e3}). To find the consequences of such a
transformation, it is useful to classify the interactions in terms of
a global $[\flavor]^5$\ flavor symmetry.
Each of the Yukawa terms in the superpotential breaks one of the
\flavor\ symmetries down to \suu. These are the only supersymmetric
interactions that break flavor symmetry. The three-scalar soft terms
are linear in squark and slepton fields (see Appendix~A)
and they also break the
flavor symmetry in the scalar sector down to $[\suu]^5$.
In general, the coefficients of the soft SUSY breaking terms are not
related to the Yukawa coupling constants so that the combination of
three-scalar terms and Yukawa interactions breaks the flavor symmetry
down to $[U(1)]^5$.
The squark and slepton mass terms break completely
the flavor symmetry in the scalar sector. Therefore, the
flavor transformation changes only
the coefficients of the soft SUSY breaking
terms which involve scalars. In particular, the squark mass matrices
are redefined by unitary transformations.

\section{Fermion Mass Generation}
\label{sec:mass}
\setcounter{equation}{0}

We begin the discussion of fermion mass generation by studying the
interactions responsible for the top mass.
We use the same notation for the fermions as the one in
Eqs.~(\ref{e1})-(\ref{e3}) for the corresponding chiral superfields
and we switch from two-component to four component spinors.
We denote the scalars by the symbols used for their
fermion partners with a tilde and we define ``right-handed'' scalar
fields: $\fir \equiv \tilde{\phi}^{c^{\dagger}}$, etc.
In the case of the third generation fermions, we use the conventional
notation, $t, b, \tau$, without distinguishing between weak and mass
eigenstates.

The top quark has Yukawa
interactions with the $\down$\ technifermion:
\be
C_q\bar{\down}_{\inr}t_{\inl}\fil +
C_t\bar{t}_{\inr}\down_{\inl}\tilde{\phi}^{\dagger}_{\inr}
+ \hc
\label{e6}
\ee

Both the soft SUSY breaking terms and the mass terms
in the superpotential contribute to the \fil -\fir\ mass matrix
(see Eq.~(\ref{mmfi})).
At energies lower than the masses of \fil\ and \fir\,
the exchange of these scalars gives rise to four-fermion
operators involving two quarks and two technifermions. The top mass
arises due to a four-fermion operator which requires
\fil -\fir\ mixing in the techni-scalar exchange, as
shown in \ref{fig:top}:
\be
\frac{C_qC_t}{\mfis}\left(\bar{\down}_{\inl}t_{\inr}\right)
\left(\bar{t}_{\inl}\down_{\inr}\right) + \hc~,
\label{e7}
\ee
where \mfis\ is a combination of the diagonal, \mfil, \mfir, and
off-diagonal, \mfilr, elements of the mass matrix for \fil\ and \fir,
\be
\mfis = \frac{\mfil\mfir - \mfilrs}{\mfilr}~.
\label{e8}
\ee
Applying a Fierz transformation to the operator~(\ref{e7}) we find
the top mass:
\be
m_t = \frac{C_qC_t}{4\mfis}\langle\bar{\down}\down\rangle~.
\label{e9}
\ee
According to naive dimensional analysis \cite{manohar},
the condensate is related to the technipion decay constant $v$\
by $\langle\bar{\down}\down\rangle \approx 4\pi v^3$.
In one-doublet technicolor models, $v \approx 246 \, \gev$.
The constraint on \mfis\ imposed by Eq.~(\ref{e9}) is
\be
\mfit \approx 0.5 \, \tev \times (C_qC_t)^{\frac{1}{2}}~,
\label{e10}
\ee
where we used $m_t = 176 \, \gev$ \cite{top}.
Since perturbation theory requires Yukawa couplings smaller than
$\sim 4 \pi$, this equation places an upper bound on the masses
of the $\tilde{\phi}$\ scalars in the absence of fine-tuning in
the mass matrix for \fil\ and \fir. We will assume
$\mfit \sim  1 \tev$.
The technicolor corrections to the diagram shown in \ref{fig:top}
are not important since the constituent masses of the
$\tilde{\phi}$\ scalars are of the order of the current mass $\mfit$.
Note that the Yukawa interactions may be treated as a small perturbation
on the technicolor dynamics \cite{steve} because $m_t/4\pi v$\
is small.

The $b$\ quark acquires mass similarly, by coupling to the $\up$\
technifermion. Since the custodial $SU(2)_R$\ symmetry requires
$\langle\bar{\down}\down\rangle \approx \langle\bar{\up}\up\rangle$,
the only source for the large mass ratio of the $t$\ and $b$\ quarks
is the ratio of the Yukawa couplings,
\be
\frac{m_t}{m_b} = \frac{C_t}{C_b}~.
\label{e11}
\ee
The $\tau$\ mass is produced by the exchange of $\tilde{\chi}$ scalars,
so the $t$\ to $\tau$\ mass ratio depends on the scalar masses:
\be
\frac{m_t}{m_{\tau}} = \frac{C_qC_t}{C_lC_{\tau}}
\left(\frac{\mchit}{\mfit}\right)^2~,
\label{e12}
\ee
where $\mchis$\ is a combination of the elements of the mass matrix for
\chil\ and \chir\ analogous to Eq.~(\ref{e8}).

Eq.~(\ref{e10}) provides some information about the
SUSY breaking scale $\mss$.
As long as we do not refer to the high energy theory responsible for
SUSY breaking there are no theoretical constraints on the coefficients
of the soft SUSY breaking terms.
However, we assume these coefficients to have the same order of
magnitude, given by $\mss$.
The masses of the $\tilde{\phi}$ scalars indicate
\be
\mss \sim \cO(1\, \tev)~.
\label{e12a}
\ee

Note that in general the technifermion masses $m_{\phi}$\ and
$m_{\chi}$, the techniscalar mixings \mfilr\ and \mchilr\,
and the Yukawa coupling constants $C_f (f = q,t,b,l,\tau)$\
are complex numbers. However, their phases can be absorbed in the scalar
and fermion fields so that all the quantities which appear in
Eqs.~(\ref{e8})-(\ref{e12}) are real.
The phase redefinition can be done in several ways and introduces new
complex phases in other coupling constants. In the discussion of CP
violation (see Section~\ref{subsec:cp})
we will use an explicit phase convention.

\vspace{0.2cm}
Although the quarks of the first and second generations do not couple
to the technifermions, there are contributions to their masses from
interactions with gauginos and squarks. The electroweak
symmetry breaking enters in these radiative masses through the mixing
of the left-handed and right-handed squarks.
This chirality-flip mixing is produced by the exchange of the $\phi$\
technifermion, whose mass, $m_{\phi}$, is an arbitrary parameter in the
superpotential. For simplicity, we will assume $m_{\phi} \sim \mss$.

In the ``super-weak'' basis, where the quark-squark-gluino vertices are
flavor diagonal and the superpotential is given by Eq.~(\ref{e5}),
only the squarks of the third generation couple to the technifermions.
However, in this basis the left-handed and right-handed squark mass
matrices are in general non-diagonal. The off-diagonal squark masses
combined with the $\tilde{u}_{\trei_\inl}$-$\tilde{u}_{\trei_\inr}$
mixing produce chirality-flip mixings
of the $\tilde{u}_\unu$\ and $\tilde{u}_\doi$\ squarks.
The quark-squark-gluino interaction leads to the one-loop graph shown in
\ref{fig:box} which yield an effective four-fermion interaction.
When the technifermions condense, this graph makes the largest
contribution to the elements of the up quark mass matrix:
\be
\muij = \frac{\alpha_s}{\pi} m_t
\frac{m_{\phi} \mfis}{m_{\tilde{g}}^3} \fuij~,
\label{e15}
\ee
where $i,j = 1,2,3$,\
$\, \fuij$\ are functions of the squark and gluino masses
given in Appendix~B,
$\alpha_s \approx 0.1$\ is the strong coupling
constant at a scale $\sim \mss$\ and $m_{\tilde{g}}$\ is the gluino
mass. A rough estimate (see Eq.~(\ref{Afij})) gives
$|\fuij| \lae 10^{-1}$.
Hence, the \mtt\ element of the quark mass matrix is given by
Eq.~(\ref{e9}) ($\mtt \approx m_t$) while the other elements are much
smaller,
\be
\frac{|\muij|}{m_t} \leq \cO(10^{-2})~.
\label{e16}
\ee
One of the benefits of this structure of the quark mass matrix is that
analytical expressions for the quark masses may be obtained.
Diagonalizing the mass matrix we find the quark masses
to first order in $\muij/m_t$:
\bear
m_c = \left( |\muu|^2 + |\muc|^2 + |\mcu|^2 + |\mcc|^2
\right)^{\frac{1}{2}}
\nonumber \\ [0.2cm]
m_u = \frac{ \left|\muu\mcc - \muc\mcu\right| }
{\left( |\muu|^2 + |\muc|^2 + |\mcu|^2 + |\mcc|^2
\right)^{\frac{1}{2}}}~.
\label{e17}
\eear

Since the supersymmetric part of the Lagrangian has a flavor
\fla symmetry with respect to the first and second generations,
the super-weak basis is defined up to such a transformation.
Therefore, there is a super-weak basis in which the $(1,2)$ elements of
the left-handed and right-handed squark mass matrices, $\mlq$
and $\mru$, vanish. The complex phases of the other non-diagonal
elements can be absorbed in the definition of the squark fields,
such that the squark mass matrices are real and symmetric. We will
assume the $(2,3)$ elements and the diagonal elements to be of order
\ms. As we will discuss in Section~\ref{subsec:fcnc},
the constraints from FCNC require small values of the $(1,3)$ elements:
\bear
\eql \equiv \frac{\mlut}{\mlct} \lae \cO(10^{-1})
\nonumber \\ [2mm]
\eur \equiv \frac{\mrut}{\mrct} \lae \cO(10^{-1})
\label{epsilon}
\eear
This structure of the squark mass matrices allow us to obtain the
unitary matrices $U_\inl^q$ and $U_\inr^u$ which transform the squark
fields from the super-weak basis to the mass eigenstate basis;
to first order in $\eql$,
\be
U_\inl^q = \left(\ba{rcl} 1 \; & \cO(\eql) & \cO(\eql) \\ [1mm]
	\cO(\eql) & \cos\theta_\inl & - \sin\theta_\inl \\ [1mm]
	\cO(\eql) & \sin\theta_\inl & \cos\theta_\inl   \\
	\ea\right)~,
\label{unitar}
\ee
where
\be
\cos^2\theta_\inl = \frac{1}{2} \left\{ 1 +
\left|\mlcc - \mltt\right|\left[ \left(\mlcc - \mltt\right)^2 + 4
\mlcts
\right]^{-1/2} \right\}
\label{[Acos}
\ee
and similar relations hold for $U^u_\inr$. Eqs.~(\ref{unitar}),
(\ref{Afij}) and (\ref{e15}) yield the following structure for the
up quark mass matrix:
\be
\mU = m_t \left(\ba{rcl}
\cO(\eql\eur \bet) & \cO(\eql \bet) & \cO(\eql \bet)\\ [1mm]
\cO(\eur \bet) & \; \bet   & \cO(\bet) \\ [1mm]
\cO(\eur \bet) & \cO(\bet) & \; \; 1    \\ \ea\right)~,
\label{matrix}
\ee
where
\be
\bet = \frac{\alpha_s}{\pi}
\frac{m_{\phi} \mfis}{m_{\tilde{g}}^3} \fucc~.
\label{beta}
\ee
The up and charm masses can be estimated by combining Eqs.~(\ref{e17})
and (\ref{matrix}):
\bear
 \frac{m_c}{m_t} & \sim & \bet \\
 \frac{m_u}{m_c} & \sim & \cO(\eql\eur)
\label{ratio}
\eear
These are realistic predictions \cite{data}, provided
$\eql\eur \sim 10^{-2}$ (see Eq.~(\ref{epsilon})) and
$\bet \sim 10^{-2}$. Note that the quark masses computed here are
at a scale of order $\mss$ and are smaller by a factor $\sim 2$
than the masses at a scale of 1 GeV \cite{kagan}.
The quark mass matrix \mU\ is diagonalized by unitary matrices:
\be
\vul^\dagger \mU \vur = {\rm diag} (m_u,m_c,m_t)
\ee
where, to first order in $\bet$ and $\eql$,
\be
\vul = \left(\ba{rcl} 1 \; & \cO(\eql) & \cO(\eql \bet)\\ [1mm]
	\cO(\eql) & \; 1 & \cO(\bet)\\ [1mm]
	\cO(\eql \bet) & \cO(\bet) & \; \; 1 \\ \ea\right)~;
\label{qrot}
\ee
\vur\ has the same structure, with \eur\ instead of \eql.

The elements of the down quark mass matrix are given by Eq.~(\ref{e15})
with $m_t$ replaced by $m_b$ and with different functions $f^d_{ij}$.
The comparatively large strange to bottom mass ratio,
$\beta^d$, given by Eq.~(\ref{beta}) with $f_{22}^u$ replaced by
$f_{22}^d$,
requires the ratio $M_{\tilde{\phi}}/m_{\tilde{g}}$ to have a rather
large value $\sim 3$ and $f_{22}^d$ to be close to its upper bound $\sim
10^{-1}$. The difference between the $m_s/m_b$ and $m_c/m_t$ ratios
is due to the different squark mass matrices which contribute to
$f_{22}^u$, respectively $f_{22}^d$.

The FCNC constraints on the down squark sector are stronger (see
Section~\ref{subsec:fcnc}), giving an upper bound
\be
\eql \edr \lae \cO(10^{-3})~,
\label{epsilond}
\ee
where
\be
\edr \equiv \frac{\mrdb}{\mrsb}~.
\label{epsilondp}
\ee
This makes the contribution of the one-loop graph of
\ref{fig:box} (with up-type quarks and squarks replaced by
down-type ones) to the down mass, $m_d$, very small.
However, there are other contributions from two-loop graphs in which the
$\tilde{d_\unu}_\inl$-$\tilde{d_\unu}_\inr$ mixing is mediated by
three-scalar interactions and techni-gluino exchange (see
\ref{fig:down}). The coefficient of the
$\tilde{d_\unu}_\inl$-$\tilde{d_\unu}_\inr$ mass term
produced is of order
$\mu_{d_\unu} \mu_{q_\unu}/(4\pi)^2$ where $\mu_{d_\unu}$ and
$\mu_{q_\unu}$ are the mass coefficients of the
$\tilde{d}\tilde{\phi}\tilde{\up}$ soft SUSY breaking terms (see
Eq.~(\ref{trilinear})). The down quark mass produced
is large enough provided $\mu_{d_\unu} \mu_{q_\unu} \sim \ms$.

The down quark mass matrix is diagonalized by unitary matrices,
\vdl\ and \vdr, with
the same structure as \vul. Therefore, the CKM matrix
\be
V_{KM} = \vul^\dagger \vdl
\label{km}
\ee
has also the structure shown in Eq.~(\ref{qrot}), with elements
$V_{us} \sim \cO(10^{-1})$, $V_{cb}\sim \cO(10^{-2})$ and
$V_{cb} \sim \cO(10^{-3})$.

In the case of charged leptons, the elements of the mass matrix are
given by one-loop graphs similar to the one in \ref{fig:box}
with the gluino, the $\phi$ technifermion and the squarks
replaced, respectively, by a zino or photino, a $\chi$ technifermion
and sleptons.
The $m_\mu/m_\tau$ ratio differ from $m_c/m_t$ by a factor
$\sim (\alpha_2/\alpha_s) (\mchis/\mfis)$ where $\alpha_2$
is the weak coupling constant at a scale $\sim \mss$.
Since $m_t/m_\tau \approx 100$, Eq.~(\ref{e12}) indicates a large
$\mchis/\mfis$ ratio; thus, the ratio
$m_\mu/m_\tau \approx \frac{1}{15}$ can be readily obtained.
If the charged slepton mass matrices and
the squark mass matrices have a similar structure, as it is
suggested by the constraints from $\mu \rightarrow e\gamma$
\cite{masiero}, then the electron mass is predicted to be
two orders of magnitude smaller than the muon mass.
The neutrinos remain massless because we did not introduce
right-handed spinors.

In conclusion, the mass hierarchy between the fermion
generations is established.
It is remarkable that the SUSY-TC model is able to reproduce
the complicated pattern of fermion masses with only few assumptions
about the soft SUSY breaking terms and the parameters in the
superpotential.

\section{Flavor-Changing Neutral Currents}
\label{sec:fcnc}
\setcounter{equation}{0}

The measurements of FCNC effects impose severe
constraints on ETC models and on SUSY models. Therefore, FCNC
represent an important test for a SUSY-TC model. In this section
we discuss the FCNC in our model, concentrating on the quark sector.

\subsection{Neutral meson mixing}
\label{subsec:fcnc}

As we showed in Section 2, in the super-weak basis only the $b$\ and
$t$\ quarks couple to the technifermions.
However, quark mixings are produced at the one loop level and, as
a result, the quarks of the first and second generations
in the mass eigenstate basis have Yukawa
interactions with the technicolored fields proportional to the
small mixing angles of the third generation.
Thus, box diagrams with techniscalars and
technifermions in the internal lines contribute to
$K-\bar{K}$\ and $B-\bar{B}$\ mixing. Nevertheless, these contributions
are suppressed by a factor of order
$\ms/M_W^2$\ with respect to the SM amplitudes
and can be ignored. Other contributions to
the $\Delta S = 2$\ or $\Delta B = 2$\ amplitudes involving
technicolored fields are given by dimension-12 operators and are much
smaller.

Larger FCNC are produced due to the techni-singlet sparticles.
In generic SUSY models \cite{masiero},
the quark and squark mass matrices are diagonalized by different
transformations.
Therefore, the quark-squark-gaugino vertices are flavor
non-diagonal in the mass eigenstate basis and give rise to FCNC.
It is convenient to compute the FCNC effects in
the super-KM basis \cite{dugan}, where the quark-squark-gluino
vertices are flavor diagonal, using the mass insertion
approximation.
This procedure was used extensively \cite{masiero} to put bounds
on the off-diagonal elements of the left-handed and right-handed
squark mass matrices and
on the chirality-flip mixings of squarks belonging to different
generations.
The tightest bounds are on down-type squark mixings and
come from gluino one-loop diagrams contributing to
$b \rightarrow s\gamma$\ and to $K-\bar{K}$\ and $B-\bar{B}$\ mixing.

In our model, the chirality-flip mixing of the down-type squarks
arises due to the $\phi$ technifermion exchange diagram shown in
\ref{fig:scalar}. At energies below $m_{\phi}$,
the effect of $\phi$\ exchange may be approximated by local
operators of dimension-five. The rules of naive dimensional analysis
show that this is a good approximation if
\be
m_{\phi} > v \frac{C_qC_b}{4\pi}~.
\label{mfi}
\ee
When the technifermions condense, the mixing of the scalar-bottoms
in the super-weak basis is given by
\be
\mlrb = - \frac{C_qC_b}{2m_{\phi}}\langle\bar{\up}\up\rangle~.
\label{blr}
\ee
Combining Eqs.~(\ref{e9}), (\ref{e11}) and (\ref{blr}) gives
\be
\left|\mlrb\right| = 2m_b\frac{\mfis}{m_{\phi}}~,
\label{bmix}
\ee
which is a small mixing: $\mlrb/\ms \sim 10^{-2}$.
The chirality-flip mixings of the down-type
squarks belonging to different generations are composed of a $(3,i)$
or $(i,3)\; (i = 1,2)$ element of \mlq\ or $\mrd$
and the $\tilde{d}_{\trei_\inl}$-$\tilde{d}_{\trei_\inr}$ mixing given
by Eq.~(\ref{bmix}).
The chirality-flip mixing produced by techni-gluino exchange and
trilinear scalar interactions, as in \ref{fig:down},
is also small:
\be
\frac{ \left|\mlrij\right| }{\ms} \sim \cO(10^{-2})~,
\ee
where $i,j = 1,2,3$.
In the super-KM basis, the $\tilde{d}_\unu$ and $\tilde{d}_\doi$ squarks
couple to the technifermions with Yukawa couplings suppressed by the
$b$-$d$ and $b$-$s$ quark mixing angles, respectively.
This also produces very small chirality-flip squark mixings.
We conclude that   the stringent bounds from $K-\bar{K}$, $B-\bar{B}$\
mixing and $b \rightarrow s\gamma$\ on the chirality-flip
squark mixing \cite{masiero} are naturally satisfied in our model.

There are, however, important constraints on the \mlij\
and \mrdij\ $(i \neq j)$ mixings due to the gluino box
diagrams contributing to the $K-\bar{K}$\ mass difference
\cite{masiero}
\bear
\frac{ \left|\mldsp\right| }{\ms}, \;
\frac{ \left|\mrdsp\right| }{\ms} \leq \cO(10^{-1})
\frac{\mss}{1 \tev} \nonumber\\
\left| \frac{\mldsp}{\ms} \frac{\mrdsp}{\ms}
\right|^{1/2} \leq \cO(10^{-2})\frac{\mss}{1 \tev}
\label{e3.1a}
\eear
and to the $B-\bar{B}$ mass difference
\be
\frac{ \left|\mldbp\right| }{\ms}, \;
\frac{ \left|\mrdbp\right| }{\ms} \leq \cO(10^{-1})
\frac{\mss}{1 \tev}
\label{e3.1b}
\ee
where the primed matrix elements refer to the super-KM basis.

The constraints on chirality-conserving mixing between the squarks of
the second and third generations are loose. The
$B_s-\bar{B_s}$\ mass difference is expected to be larger by an order
of magnitude than the $B-\bar{B}$\ mass difference \cite{rosner}.
Combined with Eq.~(\ref{e3.1b}), this leads to
\be
\frac{ \left|\mlsbp\right| }{\ms}, \;
\frac{ \left|\mrsbp\right| }{\ms} \leq \cO(1)
\frac{\mss}{1 \tev}~.
\label{e3.2}
\ee

The up squark mass matrices are also constrained; the gluino box
diagrams contributing to the $D-\bar{D}$\ mass difference
\cite{masiero} require
\be
\frac{ \left|\mlucp\right| }{\ms},\;
\frac{ \left|\mrucp\right| }{\ms} \leq \cO(10^{-1})
\frac{\mss}{1 \tev}~.
\label{e3.2p}
\ee
The wino box diagrams contributing to the $K-\bar{K}$\ mass difference
give an upper limit for the squark mixing which is larger than the
bound in Eq.~(\ref{e3.2p}).

The small values of the ratios in Eqs.~(\ref{e3.1a}), (\ref{e3.1b})
and (\ref{e3.2p}) are unnatural. In general one expects these ratios
to be of order one \cite{hall}.
A possible solution to this problem might be the existence of gauged
horizontal symmetries \cite{leigh}.
Note that the bounds given by Eqs.~(\ref{e3.1a})-(\ref{e3.2p})
are $\sim \mss$\ which implies looser bounds in our model than in
the SSM where the SUSY breaking scale is likely to be below 1 TeV.

The $(1,2)$ elements of the down squark mass matrices in the super-KM
basis are related to the ones in the super-weak basis by:
\bear
\left|\mldsp\right| \approx \eql \left|\mluu - \mlcc\right| \nonumber\\
\left|\mrdsp\right| \approx \edr \left|\mrdd - \mrss\right|
\label{rel}
\eear
where we kept only the leading terms in $\beta^d$, \eql\ and \edr.
The relation between the Cabbibo angle and \mluc, given by
Eq.~(\ref{qrot}), requires $\eql \sim \cO(10^{-1})$.
Therefore, \edr\ is strongly constrained by Eq.~(\ref{e3.1a}). In order
to avoid excessive fine-tuning of \mrdb, we assume a reasonably small
value of
\be
\left[\frac{\mluu - \mlcc}{\ms}\right]
\left[\frac{\mrdd - \mrss}{\ms}\right] \sim \cO(10^{-1})~.
\ee
In this case, the bound in Eq.~(\ref{e3.1a}) is saturated if
$\edr \sim \cO(10^{-2})$. This is also a sufficient condition
for satisfying the constraints on the $(1,3)$ elements given by
Eq.~(\ref{e3.1b}).

In the up squark sector, the relation between the super-KM and the
super-weak basis is given by Eq.~(\ref{rel}), with the upper index
$d$ replaced by $u$. The inequalities~(\ref{e3.2p}) are satisfied
if $\eur, \eql \lae \cO(10^{-1})$.

It is interesting that the
bounds on the $(1,2)$ elements of the squark mass matrices in
the super-KM basis correspond to bounds on the $(1,3)$ elements in the
super-weak basis.

\subsection {CP violation}
\label{subsec:cp}

To keep track of the relative complex phases of the coupling constants
relevant for CP violation we will adopt the phase convention described
below. The masses $m_{\phi}$\ and $m_{\chi}$\ become real by absorbing
their phases in the fermion fields $\phi_{\inl}$\ and $\chi_{\inl}$.
The phases of the off-diagonal mass terms \mfilr\ and \mchilr\
are absorbed in the scalar fields  \chil\ and \fil. These redefinitions
introduce new phases in the $\phi_{\inl}$-$\fil$-gaugino and
$\chi_{\inl}$-$\chil$-gaugino interactions and also in the Yukawa
couplings from the superpotential.  The phases of the $C_q$\
Yukawa coupling from the quark
and squark sectors are now different and they are absorbed in the
fermion $\Upsilon$, respectively scalar $\tilde{\Upsilon}$\ doublets,
while a new phase appears in the $\Upsilon$-$\tilde{\Upsilon}$-gaugino
interactions. A redefinition of the $u_\trei^c, \, d_\trei^c$ and
$e_\trei^c$
superfields yields $C_t, \,C_b$\ and $C_{\tau}$\ real. The $C_l$\
coupling constant has also different phases in the lepton and slepton
vertices; these are absorbed by the $l_{\trei}$\ and
$\tilde{l}_{\trei}$\ fields leading to a phase in the
$l_{\trei}$-$\tilde{l}_{\trei}$-gaugino interactions.

At this stage, the only complex coupling constants left are in the
soft SUSY breaking terms and in the fermion-scalar-gaugino interactions
mentioned above.
The gaugino masses are in general complex.  When the gaugino field
is redefined to have real mass, a new phase, \dg, is introduced in the
quark-squark-gaugino vertices. If an internal gaugino line connects
quarks of the same chirality, the phases introduced in the two vertices
cancel each other. Nevertheless, the complex phases of the gaugino
masses are relevant when the gaugino propagator connects quarks of
different chiralities, as it is the case in the diagram shown in
\ref{fig:box}.
Thus, there are contributions linear in \dg\ to the neutron dipole
moment (NDM) from the one-loop diagram shown in \ref{fig:box}
with an external photon line attached to one of the internal
lines. These are similar with the SSM contributions to the NDM
\cite{ndm}. The experimental limit on the NDM \cite{data} requires
$\dg \lae 10^{-2}$.
Also, there are corrections of order \dg\ to the phases
of all the elements of a quark mass matrix except the (3,3) one.
However, the CKM elements can be expressed in terms of quark mass
ratios such that the phase \dg\ is largely canceled out.
Therefore, we will ignore \dg\ in the discussion of the phases in the
quark mass matrices.

A squark mass matrix is hermitian and has three complex phases.
However, in the super-weak basis, the $(1,2)$\ elements of the squark
mass matrices vanish and, therefore, there are only two phases left.
These can be absorbed in the squark fields of the first two generations.
The coupling constants of the quark-squark-gaugino vertices are kept
real by including the same phases in the definition of the corresponding
quark fields.

The result of the above phase convention  is that there is no
contribution from the squark mixings to the phase of the diagram shown
in \ref{fig:box}. Thus, the leading contributions to the quark mass
matrices (see Eqs.~(\ref{e9}) and (\ref{e15})) are real.

This result has interesting consequences. The origin of CP violation
should be rather different than the one in the SM since the CKM
matrix is approximately real. Also, real quark mass matrices
are relevant for the strong CP problem. The strong CP parameter
$\bar{\theta}$\ receives in this case no contribution from the
quark mass matrix \cite{cheng} while the QCD contributions can be small
enough if CP is spontaneously broken. However, the complex phases
of the quark mass matrices that we describe below give
corrections to $\bar{\theta}$\ much larger than the experimental limit
of $10^{-9}$, such that the strong CP problem persists.

\newcommand{\Mds}{\mbox{$ M_{ds}^2$}}
\newcommand{\dds}{\mbox{$ \Delta_{ds}$}}
\newcommand{\im}{\mbox{$ \rm Im$}}
\newcommand{\re}{\mbox{$ \rm Re$}}
Complex phases in the quark mass matrices come from additional
loops involving trilinear scalar interactions or
gaugino-technifermion-techniscalar interactions.
We will denote generically the complex phases of the coupling constants
of these interactions by $\De$. The two-loop corrections to the
(2,2), (2,3), (3,2) and (3,3) elements of the quark mass matrices
are small; the typical size of the phases of these elements
is $\sim 10^{-2} \De$. The other elements are smaller
(see Eq.~(\ref{qrot})), due to the structure of the squark mass
matrices.
Thus, the two-loop contributions to the imaginary parts of these
elements are larger;
writing ${\rm Arg} \left(m_{ij}^{u,d}\right) = \delta_{ij}^{u,d}$,
we estimate:
\bear
\delta_{11}^u, \delta_{11}^d, \delta_{12}^d, \delta_{21}^d,
\delta_{13}^d, \delta_{31}^d \, \sim \De \nonumber \\
\delta_{12}^u, \delta_{21}^u, \delta_{13}^u, \delta_{31}^u \,
\sim 10^{-1} \De
\label{phase}
\eear

To see the effect of these phases, we consider the Wolfenstein
parametrization of the CKM matrix \cite{wolfenstein}, keeping only the
first non-vanishing terms of the expansion in $\lambda = \sin\theta_c$,
where $\theta_c$ is the Cabbibo angle:
\be
V_{KM} = \left(\ba{rcl} 1 \;\;\;\;
& \lambda & A\lambda^3(\rho - i\eta)\\
- \lambda \;\;\; & 1 & \;\; A\lambda^2 \\
A\lambda^3(1 - \rho - i\eta) & - A\lambda^2 & \;\;\; 1 \\
\ea\right)~.
\label{wol}
\ee
Computing $V_{KM}$, as discussed in Section~\ref{sec:mass}, to first
order in $\bet, \beta^d, \eql$, gives:
\bear
\lambda &=& \left| \frac{\mds}{\mSs} - \frac{\muc}{\mcc}
\right| \sim \cO(\eql) \nonumber \\
\vspace{1mm}
A\lambda^2 &=& \left| \frac{\msb}{\mbb} - \frac{\mct}{\mtt}
\right| \sim \cO(\beta^d)
\eear
and more complicated expressions for $\eta$ and $\rho$, which, together
with the estimated phases of the quark mass terms, indicate
\bear
& \eta \sim \cO(10^{-1} \De) & \nonumber \\
& \left| \rho(1 - \rho) \right| \sim \cO(1) &
\label{rho}
\eear

The measurements of CP asymmetry in semileptonic decays of $K_L$
show \cite{epsilon}
\be
\frac{\im M_{12}}{\re M_{12}} \approx 6.5 \times 10^{-3}~,
\label{e3.4}
\ee
where $M_{12}$\ is the off-diagonal element of the $K - \bar{K}$\
mass matrix.
If the phase from the CKM matrix is solely responsible for CP violation,
as it is in the SM, then Eq.~(\ref{e3.4}) requires \cite{rosner}
\bear
0.2 \lae \eta_{\SM} \lae 0.6 \nonumber \\
\left| \rho_{\SM} \right| \lae 0.4
\label{eta}
\eear
Comparing Eqs.~(\ref{rho}) and (\ref{eta}), we conclude that the CP
violation provided by the CKM matrix is not sufficient in our model;
hence, the bulk of CP violation in $K - \bar{K}$\ mixing is due to
SUSY-box diagrams. The relevant phases are those of the \mldsp\ and
\mrdsp\ squark mixing in the super-KM basis.
We will denote generically these squark mixings
and their phases by \Mds\ and \dds\ respectively.
These appear in the gluino box diagrams:
\be
\im M_{12} \approx \tan(2 \dds) \re (M_{12})_{gluino}~.
\label{e3.5}
\ee
Eq.~(\ref{e3.1a}) can be written as
\be
\frac{\re (M_{12})_{gluino}}{\re M_{12}} \sim
\left( 10^2 \frac{ \left|\Mds\right| }{\ms} \right)^2~,
\label{e3.6}
\ee
which combined with Eqs.~(\ref{e3.4}) and (\ref{e3.5}) gives
\be
\left(10^2 \frac{ \left|\Mds\right| }{\ms}\right)^{2}
\tan(2 \dds) \sim \cO(10^{-2})~.
\label{e3.7}
\ee
If the mass ratios in Eq.~(\ref{e3.1a}) are
close to their limits, then Eq.~(\ref{e3.7}) indicates the size of the
phases of the squark mass mixings:
\be
\dds \sim \cO(10^{-2})~.
\label{e3.8}
\ee
An explicit computation of the squark mass matrices in the super-KM basis,
involving unitary transformations with the \vdl\ and \vdr\ matrices on
the squark mass matrices in the super-weak basis, shows
$\dds \sim \delta_{12}^d$.
Comparing, then, Eqs.~(\ref{phase}) and (\ref{e3.8}), we obtain the size
of the phases of the three-scalar interactions:
\be
\De \sim \cO(10^{-2})~.
\label{del}
\ee
Such small phases might arise
naturally if there is spontaneous CP violation \cite{spcp}.

The SM predicts large CP asymmetries in $B$\ meson decays \cite{sanda}
because the phase in the CKM matrix is $\cO(1)$ (see Eq.(\ref{eta})).
Eqs.~(\ref{rho}) and (\ref{del}) show that the situation is totally
different in our model: the phases responsible for CP violation in $B$
meson decays are of order
\be
\eta \sim \cO(10^{-3})~.
\label{etatc}
\ee
Note that the one-loop diagrams with sparticles in
internal lines give small contributions to the $B$ decays. In
particular, the $B - \bar{B}$ mixing amplitude given by SUSY box
diagrams is small because of the severe bounds on \eql\ and \edr\ from
$K - \bar{K}$ mixing (see Eqs.~(\ref{e3.1a}), (\ref{e3.1b}) and
(\ref{epsilond})).
Hence, the mechanism for CP violation in our model is the same as in the
SM, but the effects are smaller by a factor $\eta_{\SM}/\eta \sim
10^2$. For example, we estimate the size of CP asymmetry in
$B \rightarrow \Psi K_S$\ \cite{sanda} to be
\be
- \im\left[ \frac{q}{p}
\frac{A(\bar{B}^0 \rightarrow \Psi K_S)}
{A\left(B^0 \rightarrow \Psi K_S\right)}\right]
\approx \frac{2\eta}{1- \rho} \sim \cO(10^{-3})~.
\label{e3.10}
\ee
where $q/p$ is the $B - \bar{B}$ mixing parameter.
This value is two orders of magnitude smaller than the resolution
of the proposed experiments \cite{bexp} on CP violation in $B$\ decays.

Another consequence of Eq.~(\ref{etatc}) is a small direct CP
violation in $K_L \rightarrow \pi\pi$\ decays.
To see this, note that in the SM the $\varepsilon^{\prime}/\varepsilon$\
parameter is proportional to $\eta_{\SM}$ \cite{buras}.
Therefore, a scaling of the SM result gives
$\varepsilon^{\prime}/\varepsilon
\sim \cO(10^{-6})$ in the SUSY-TC model. The contributions from SUSY
penguin diagrams do not exceed this value.
Such a small value is inconsistent with the result of the CERN NA31
experiment ($\varepsilon^{\prime}/\varepsilon = (23 \pm 7) \times
10^{-4}$)
\cite{cern} but is consistent with the result of the Fermilab E731
experiment ($\varepsilon^{\prime}/\varepsilon = (7.4 \pm 6.0) \times
10^{-4}$) \cite{fermilab}.

\section{Conclusions}

We have proposed a supersymmetric one-doublet technicolor model
with the superpotential containing Yukawa couplings of the quarks and
leptons of the third generation with a technifermion and a
$SU(2)_W$-singlet techniscalar.
These interactions give rise to masses for the fermions of the third
generation. The fermions of the other generations have radiative masses
such that a correct mass hierarchy arises. However, the model offer no
insight into the origin of the large top to bottom mass ratio, given by
a ratio of Yukawa coupling constants. In order to
obtain a realistic top mass, the SUSY breaking scale should be
$\sim 1$ TeV. In the low energy SUSY theory, the sparticle masses are
not determined; consequently, it is not possible to make more precise
predictions for the fermion mass matrices.

The contributions of the technicolored particles to $K-\bar{K}$\ and
$B-\bar{B}$\ mixing are small. Comparing with the SSM,
the amount of fine-tuning in the squark mass matrices required to avoid
large FCNC is slightly reduced in our model.

With an appropriate redefinition, the complex phase in the CKM matrix is
$\cO(10^{-3})$. The main contributions to CP violation in $K - \bar{K}$
mixing come from gluino box diagrams.
The mechanisms for CP violation in $B$ meson decays and
for direct CP violation in $K_{\inl}$\ decays are the same as in the SM.
The CP asymmetries in $B$ decays are smaller by two orders of magnitude
than the asymmetries predicted in the SM and will not be
detected at the proposed $B$-factories. Also, the CP asymmetry in
$\Delta S = 1$ transitions is tiny.

To decide whether the model is viable it is necessary to explore many
other phenomenological issues: electroweak precision measurements,
FCNC in the lepton sector, constraints from cosmology, etc.
Also, it is interesting to study how this SUSY-TC model fits
into a high energy theory, such as grand unification or supergravity.

\section*{Acknowledgements}

I am grateful to Sekhar Chivukula for many valuable observations and
helpful discussions, and for helping me to understand the flavor
structure of the model presented here.
I would like to thank Elizabeth Simmons and
John Terning for a critical reading of the manuscript and for
useful suggestions. I thank Bhashyam Balaji, Serban Dobrescu,
Benjamin Grinstein, Dimitris Kominis and Ryan Rohm for useful
comments and discussions. Also, I thank Alexander Kagan for bringing
Refs.~\cite{kagan1, kagan2} to my attention and for useful comments.
{\em This work was supported in part by the National Science
Foundation under grant PHY-9057173, and by the Department of Energy
under grant DE-FG02-91ER40676.}

\renewcommand{\thesection}{Alph{section}}
\setcounter{section}{0}
\section*{Appendix A}
\renewcommand{\theequation}{A.\arabic{equation}}
\setcounter{equation}{0}

This Appendix presents the soft SUSY breaking terms.
We use the notation described at the beginning of
Section~\ref{sec:mass}.

The techniscalar mass terms can be written
\be
{\cal L}_{\doi s} = M^2_{\tilde{\Upsilon}} \tilde{\Upsilon}^\dagger
\tilde{\Upsilon} +
M^2_{\tilde{p}_\inl} \tilde{\up}_\inl^\dagger
\tilde{\up}_\inl +
M^2_{\tilde{m}_\inl} \tilde{\down}_\inl^\dagger
\tilde{\down}_\inl +
M^{\prime 2}_{\tilde{\phi}} \tilde{\phi}^\dagger \tilde{\phi} +
M^{\prime 2}_{\tilde{\chi}} \tilde{\chi}^\dagger \tilde{\chi}~,
\label{tsmass}
\ee
where $M^{\prime 2}_{\tilde{\phi}}$ and $M^{\prime 2}_{\tilde{\chi}}$
are $2\times 2$ hermitian matrices,
$\tilde{\phi} \equiv (\tilde{\phi}_\inl , \tilde{\phi}_\inr)^\top$
and $\tilde{\chi} \equiv (\tilde{\chi}_\inl , \tilde{\chi}_\inr)^\top$.
The mass terms in the superpotential also contribute to the
$SU(2)_W$-singlet techniscalar mass matrices:
\be
M^2_{\tilde{\phi}} = M^{\prime 2}_{\tilde{\phi}} + m_\phi^2 \check{I}~,
\label{mmfi}
\ee
where $\check{I}$ is the $2\times 2$ unit matrix.

The squark and slepton mass terms are given by $(i,j = 1,2,3)$
\be
{\cal L}_{\doi s}^\prime = \mlij \tilde{q}^\dagger_i \tilde{q}_j +
\mruij \tilde{u}_{i_\inr}^\dagger \tilde{u}_{j_\inr} +
\mrdij \tilde{d}_{i_\inr}^\dagger \tilde{d}_{i_\inr} +
\mllij \tilde{l}^\dagger_i \tilde{l}_j +
\mreij \tilde{e}_{i_\inr}^\dagger \tilde{e}_{j_\inr}~.
\label{bilinear}
\ee
The $3\times 3$ squark and slepton mass matrices are real, symmetric
and have vanishing $(1,2)$ and $(2,1)$ elements (see
Section~\ref{sec:mass}).

The trilinear scalar terms in the super-weak basis, defined by
the choice of the superpotential in Eq.~(\ref{e5}) and of the
squark and slepton mass terms in Eq.~(\ref{bilinear}), are given by:
\bear
{\cal L}_{\trei s} &=&
\mu_{q_i} \epsilon^{\alpha\beta} \tilde{q}_{i_{\alpha}}
\tilde{\Upsilon}_{\beta} \tilde{\phi}_{\inl}
+ \mu_{u_i} \tilde{u}_{i_\inr}^\dagger \tilde{\down}_{\inl}
\tilde{\phi}^\dagger_\inr
+ \mu_{d_i} \tilde{d}_{i_\inr}^\dagger \tilde{\up}_{\inl}
\tilde{\phi}^\dagger_\inr \\ \nonumber
&&
+ \, \mu_{l_i} \epsilon^{\alpha\beta} \tilde{l}_{i_{\alpha}}
\tilde{\Upsilon}_{\beta} \tilde{\chi}_{\inl}
+ \mu_{e_i} \tilde{e}_{i_\inr}^\dagger \tilde{\up}_{\inl}
\tilde{\chi}^\dagger_\inr + \hc~,
\label{trilinear}
\eear
where $\mu_{q_i}, \mu_{u_i}, \mu_{d_i}, \mu_{l_i}, \mu_{e_i}, \,
i=1,2,3$ are mass parameters.
Although these terms are linear in squark and slepton fields,
in the super-weak basis the flavors are uniquely defined and, in
general, all the generations couple to the techniscalars.

Finally, the soft SUSY terms include majorana mass terms for all the
gauginos: bino, winos, gluinos and techni-gluinos.

\section*{Appendix B}
\renewcommand{\theequation}{B.\arabic{equation}}
\setcounter{equation}{0}

In this Appendix we derive the functions $\fuij$ which enter
in the elements of the up quark mass matrix (see Eq.~(\ref{e15})).

At energies higher than the technicolor scale $\Lambda \approx 4\pi v$\
the technifermion condensate breaks and the chirality-flip mixing
of the squarks vanishes. Hence, the integral corresponding to the
one-loop graph in \ref{fig:box} should be cut off at $\Lambda$. The
$(i,j)\; (i,j = 1,2,3)$ element of the up quark mass matrix is given by:
\bear
\muij& \!=& \! \frac{16}{3}\pi \alpha_s m_{\tilde{g}}m_{\phi}C_qC_t
\langle\bar{\down}\down\rangle \nonumber \\ \vspace{0.15cm}
       &  \times& \!
\!\int^{\Lambda}\frac{d^4 p}{(2\pi)^4}
\frac{- i}{ (p^2 - m^2_{\tilde{g}})(p^2 - m^2_{\phi}) }
\sum_{k,l=1}^3
\frac{(U_\inl^q)_{3k}^* (U_\inl^q)_{ik} (U_\inr^u)_{jl}^*
(U_\inr^u)_{3l}}
{(p^2 - \Mkl)(p^2 - \Mlr)}~,
\label{eA2}
\eear
where \Mil\ (\Mir), are the eigenvalues of the left-handed
(right-handed) up squark mass matrix, \mlq\ (\mru),
and $U_\inl^q$ ($U_\inr^u$)
is the unitary matrix which diagonalizes \mlq\ (\mru, respectively).
Integrating over
the angles, using Eq.~(\ref{e9}) and comparing Eqs.~(\ref{e15}) and
(\ref{eA2}) we find
\be
\fuij = \frac{4}{3}
\int^{ \Lambda^2/m_{\tilde{g}}^2 }_{0} dy
\frac{y}{(y + 1)(y + m^2_{\phi}/m_{\tilde{g}}^2)}
\sum_{k,l=1}^3
\frac{(U_\inl^q)_{3k}^* (U_\inl^q)_{ik} (U_\inr^u)_{jl}^*
(U_\inr^u)_{3l}}
{(y + \Mkl/m_{\tilde{g}}^2)(y + \Mlr/m_{\tilde{g}}^2)}~.
\label{Afij}
\ee
The unitarity of $U_\inl^q$ and $U_\inr^u$ and the structure of the
integrand in Eq.~(\ref{Afij}) indicates an upper bound
$\fuij \lae \cO(10^{-1})$.

\newcommand{\np}{Nucl.\ Phys.\ {\bf B}}
\newcommand{\pr}{Phys.\ Rev.\ }
\newcommand{\prd}{Phys.\ Rev.\ {\bf D}}
\newcommand{\prp}{Phys.\ Rep.\ }
\newcommand{\prl}{Phys.\ Rev.\ Lett.\ }
\newcommand{\pl}{Phys.\ Lett.\ {\bf B}}
\newcommand{\ptp}{Prog.\ Theor.\ Phys.\ }
\newcommand{\ap}{Ann.\ Phys.\ }
\newcommand{\intl}{Int.\ J.\ Mod.\ Phys.\ {\bf A}}

\vspace{1.5cm}
\newpage
\noindent
{\Large\bf Figure captions}
\vspace{0.5cm}

\begin{enumerate}
\renewcommand{\theenumi}{Fig.~\arabic{enumi}}
\setcounter{enumi}{0}

\item \label{fig:top}
Four-fermion interaction due to the exchange of
technicolored scalars. The cross on the scalar line denotes the
chirality-flip  mixing of the $\tilde{\phi}$\ scalars.

\item \label{fig:box} The radiative correction involving a gluino,
$\tilde{g}$, gives the leading contribution to the masses of the up,
$u_1$, and charm, $u_2$, quarks.

\item \label{fig:down} The leading contribution to $m_d$ has a
techni-gluino, $\tilde{\cal G}_{TC}$, and techniscalars in the loops.

\item \label{fig:scalar} Technifermion exchange leading to a
dimension-five operator responsible for
$\tilde{d}_{\trei_\inl}$-$\tilde{d}_{\trei_\inr}$ mixing.
The cross on the fermion line represents
the \ew singlet mass term for the $\phi$ technifermion.
\end{enumerate}

\newpage
\pagestyle{empty}
\newcommand{\fel}{10480}
\newcommand{\scl}{8214}
\newcommand{\fgl}{15200}
\newcommand{\ffl}{7600}
\newcommand{\tgl}{11896}
\newcommand{\tglj}{5948}
\newcommand{\ori}{17400}
\newcommand{\oriy}{45500}
\newcommand{\orib}{15500}
\newcommand{\orx}{25000}
\newcommand{\orid}{5000}
\input FEYNMAN
\begin{picture}(30000,53000)
\drawline\scalar[\W\REG](\ori,\oriy)[3]
\drawarrow[\LDIR\ATTIP](\pmidx,\pmidy)
\global\advance\pmidy by -2000
\put(\pmidx,\pmidy){\fil}
\drawline\fermion[\NW\REG](\pbackx,\oriy)[\ffl]
\drawarrow[\LDIR\ATTIP](\pmidx,\pmidy)
\global\advance\pmidx by 700
\global\advance\pmidy by 450
\put(\pmidx,\pmidy){$\down_{\inr}$}
\drawline\fermion[\SW\REG](\pfrontx,\oriy)[\ffl]
\drawarrow[\NE\ATTIP](\pmidx,\pmidy)
\global\advance\pmidx by 900
\global\advance\pmidy by -450
\put(\pmidx,\pmidy){$t_{\inl}$}
\drawline\scalar[\E\REG](\ori,\oriy)[3]
\drawarrow[\W\ATTIP](\pmidx,\pmidy)
\global\advance\pmidy by -2000
\put(\pmidx,\pmidy){\fir}
\drawline\fermion[\NE\REG](\pbackx,\oriy)[\ffl]
\drawarrow[\SW\ATTIP](\pmidx,\pmidy)
\global\advance\pmidx by -1800
\global\advance\pmidy by 450
\put(\pmidx,\pmidy){$\down_{\inl}$}
\drawline\fermion[\SE\REG](\pfrontx,\oriy)[\ffl]
\drawarrow[\LDIR\ATTIP](\pmidx,\pmidy)
\global\advance\pmidx by -1800
\global\advance\pmidy by -450
\put(\pmidx,\pmidy){$t_{\inr}$}
\THICKLINES
\drawline\fermion[\SE\REG](\ori,\oriy)[1420]
\drawline\fermion[\SW\REG](\ori,\oriy)[1420]
\drawline\fermion[\NE\REG](\ori,\oriy)[1420]
\drawline\fermion[\NW\REG](\ori,\oriy)[1420]
\THINLINES
\global\advance\pfronty by -11000
\put(\pbackx,\pfronty){\ref{fig:top}}
%
\THICKLINES
\drawline\fermion[\W\REG](\orx,\orib)[\tgl]
\drawarrow[\E\ATTIP](\pmidx,\pmidy)
\THINLINES
\global\advance\pmidy by 1000
\put(\pmidx,\pmidy){$\tilde{g}$}
\drawline\fermion[\NW\REG](\pbackx,\orib)[\ffl]
\drawarrow[\SE\ATTIP](\pmidx,\pmidy)
\global\advance\pmidx by 700
\global\advance\pmidy by 450
\put(\pmidx,\pmidy){$u_{i_{\inl}}$}
\drawline\scalar[\S\REG](\pfrontx,\pfronty)[3]
\drawarrow[\S\ATTIP](\pmidx,\pmidy)
\global\advance\pmidx by -2000
\global\advance\pmidy by -300
\put(\pmidx,\pmidy){$\tilde{u}_{i_{\inl}}$}
\THICKLINES
\drawline\fermion[\SE\REG](\pbackx,\pbacky)[1500]
\drawline\fermion[\SW\REG](\pfrontx,\pfronty)[1500]
\drawline\fermion[\NE\REG](\pfrontx,\pfronty)[1500]
\drawline\fermion[\NW\REG](\pfrontx,\pfronty)[1500]
\THINLINES
\drawline\scalar[\S\REG](\pfrontx,\pfronty)[3]
\drawarrow[\S\ATTIP](\pmidx,\pmidy)
\global\advance\pmidx by -2000
\global\advance\pmidy by -300
\put(\pmidx,\pmidy){$\tilde{u}_{\trei_\inl}$}
\drawline\fermion[\SW\REG](\pbackx,\pbacky)[\ffl]
\drawarrow[\SW\ATTIP](\pmidx,\pmidy)
\global\advance\pmidx by 900
\global\advance\pmidy by -450
\put(\pmidx,\pmidy){$\down_{\inr}$}
\drawline\fermion[\E\REG](\pfrontx,\pfronty)[\tglj]
\drawarrow[\W\ATTIP](\pmidx,\pmidy)
\global\advance\pmidy by -1800
\put(\pmidx,\pmidy){$\phi_{\inl}$}
\THICKLINES
\drawline\fermion[\SE\REG](\pbackx,\pbacky)[1420]
\drawline\fermion[\SW\REG](\pfrontx,\pfronty)[1420]
\drawline\fermion[\NE\REG](\pfrontx,\pfronty)[1420]
\drawline\fermion[\NW\REG](\pfrontx,\pfronty)[1420]
\THINLINES
\global\advance\pfronty by -10000
\put(\pbackx,\pfronty){\ref{fig:box}}
\global\advance\pfronty by 10000
\drawline\fermion[\E\REG](\pfrontx,\pfronty)[\tglj]
\drawarrow[\W\ATTIP](\pmidx,\pmidy)
\global\advance\pmidy by -1800
\put(\pmidx,\pmidy){$\phi_{\inr}$}
\drawline\fermion[\SE\REG](\pbackx,\pbacky)[\ffl]
\drawarrow[\NW\ATTIP](\pmidx,\pmidy)
\global\advance\pmidx by -1800
\global\advance\pmidy by -450
\put(\pmidx,\pmidy){$\down_{\inl}$}
\drawline\scalar[\N\REG](\pfrontx,\pfronty)[3]
\drawarrow[\N\ATTIP](\pmidx,\pmidy)
\global\advance\pmidx by 1300
\global\advance\pmidy by -300
\put(\pmidx,\pmidy){$\tilde{u}_{\trei_\inr}$}
\THICKLINES
\drawline\fermion[\SE\REG](\pbackx,\pbacky)[1500]
\drawline\fermion[\SW\REG](\pfrontx,\pfronty)[1500]
\drawline\fermion[\NE\REG](\pfrontx,\pfronty)[1500]
\drawline\fermion[\NW\REG](\pfrontx,\pfronty)[1500]
\THINLINES
\drawline\scalar[\N\REG](\pfrontx,\pfronty)[3]
\drawarrow[\N\ATTIP](\pmidx,\pmidy)
\global\advance\pmidx by 1300
\global\advance\pmidy by -300
\put(\pmidx,\pmidy){$\tilde{u}_{j_{\inr}}$}
\drawline\fermion[\NE\REG](\pbackx,\orib)[\ffl]
\drawarrow[\NE\ATTIP](\pmidx,\pmidy)
\global\advance\pmidx by -1800
\global\advance\pmidy by 450
\put(\pmidx,\pmidy){$u_{j_{\inr}}$}
\end{picture}
\vfil
\newpage
\pagestyle{empty}
\begin{picture}(30000,53000)
\THICKLINES
\drawline\fermion[\W\REG](\orx,\oriy)[\tgl]
\drawarrow[\E\ATTIP](\pmidx,\pmidy)
\global\advance\pmidy by 1000
\put(\pmidx,\pmidy){$\tilde{g}$}
\THINLINES
\drawline\fermion[\NW\REG](\pbackx,\oriy)[\scl]
\drawarrow[\SE\ATTIP](\pmidx,\pmidy)
\global\advance\pmidx by 700
\global\advance\pmidy by 450
\put(\pmidx,\pmidy){$d_{\unu_\inl}$}
\drawline\scalar[\S\REG](\pfrontx,\pfronty)[3]
\drawarrow[\S\ATTIP](\pmidx,\pmidy)
\global\advance\pmidx by -2000
\global\advance\pmidy by -300
\put(\pmidx,\pmidy){$\tilde{d}_{\unu_\inl}$}
\drawline\scalar[\E\REG](\pbackx,\pbacky)[3]
\drawarrow[\W\ATTIP](\pmidx,\pmidy)
\global\advance\pmidy by -2000
\put(\pmidx,\pmidy){\fil}
\THICKLINES
\drawline\fermion[\SE\REG](\pbackx,\pbacky)[1420]
\drawline\fermion[\SW\REG](\pfrontx,\pfronty)[1420]
\drawline\fermion[\NE\REG](\pfrontx,\pfronty)[1420]
\drawline\fermion[\NW\REG](\pfrontx,\pfronty)[1420]
\THINLINES
\drawline\scalar[\E\REG](\pfrontx,\pfronty)[3]
\drawarrow[\W\ATTIP](\pmidx,\pmidy)
\global\advance\pmidy by -1800
\put(\pmidx,\pmidy){$\phi_{\inr}$}
\global\advance\pbackx by -11896
\drawline\scalar[\S\REG](\pbackx,\pbacky)[3]
\drawarrow[\S\ATTIP](\pmidx,\pmidy)
\global\advance\pmidx by -2000
\global\advance\pmidy by -300
\put(\pmidx,\pmidy){$\tilde{\up}_{\inr}$}
\drawline\fermion[\SW\REG](\pbackx,\pbacky)[\scl]
\drawarrow[\SW\ATTIP](\pmidx,\pmidy)
\global\advance\pmidx by 900
\global\advance\pmidy by -450
\put(\pmidx,\pmidy){$\up_{\inr}$}
\THICKLINES
\drawline\fermion[\E\REG](\pfrontx,\pfronty)[\tgl]
\drawarrow[\W\ATTIP](\pmidx,\pmidy)
\global\advance\pmidy by -1800
\global\advance\pmidx by -500
\put(\pmidx,\pmidy){${\tilde{\cal G}}_{TC}$}
\THINLINES
\global\advance\pmidy by -9000
\put(\pmidx,\pmidy){\ref{fig:down}}
\drawline\fermion[\SE\REG](\pbackx,\pbacky)[\scl]
\drawarrow[\NW\ATTIP](\pmidx,\pmidy)
\global\advance\pmidx by -1800
\global\advance\pmidy by -450
\put(\pmidx,\pmidy){$\up_{\inl}$}
\drawline\scalar[\N\REG](\pfrontx,\pfronty)[3]
\drawarrow[\N\ATTIP](\pmidx,\pmidy)
\global\advance\pmidx by 1300
\global\advance\pmidy by -300
\put(\pmidx,\pmidy){$\tilde{\up}_{\inl}$}
\drawline\scalar[\N\REG](\pbackx,\pbacky)[3]
\drawarrow[\N\ATTIP](\pmidx,\pmidy)
\global\advance\pmidx by 1300
\global\advance\pmidy by -300
\put(\pmidx,\pmidy){$\tilde{d}_{\unu_\inr}$}
\drawline\fermion[\NE\REG](\pbackx,\oriy)[\scl]
\drawarrow[\NE\ATTIP](\pmidx,\pmidy)
\global\advance\pmidx by -1800
\global\advance\pmidy by 450
\put(\pmidx,\pmidy){$d_{\unu_\inr}$}
%
\drawline\fermion[\W\REG](\ori,\orid)[\tglj]
\drawarrow[\LDIR\ATTIP](\pmidx,\pmidy)
\global\advance\pmidy by -1800
\put(\pmidx,\pmidy){$\phi_{\inl}$}
\drawline\fermion[\NW\REG](\pbackx,\orid)[\ffl]
\drawarrow[\LDIR\ATTIP](\pmidx,\pmidy)
\global\advance\pmidx by 700
\global\advance\pmidy by 450
\put(\pmidx,\pmidy){$\up_{\inr}$}
\drawline\scalar[\SW\REG](\pfrontx,\orid)[4]
\drawarrow[\NE\ATTIP](\pmidx,\pmidy)
\global\advance\pmidx by 1100
\global\advance\pmidy by -650
\put(\pmidx,\pmidy){$\tilde{u}_{\trei_\inl}$}
\drawline\fermion[\E\REG](\ori,\orid)[\tglj]
\drawarrow[\W\ATTIP](\pmidx,\pmidy)
\global\advance\pmidy by -1800
\put(\pmidx,\pmidy){$\phi_{\inr}$}
\drawline\fermion[\NE\REG](\pbackx,\orid)[\ffl]
\drawarrow[\SW\ATTIP](\pmidx,\pmidy)
\global\advance\pmidx by -1800
\global\advance\pmidy by 450
\put(\pmidx,\pmidy){$\up_{\inl}$}
\drawline\scalar[\SE\REG](\pfrontx,\orid)[4]
\drawarrow[\LDIR\ATTIP](\pmidx,\pmidy)
\global\advance\pmidx by -1800
\global\advance\pmidy by -650
\put(\pmidx,\pmidy){$\tilde{u}_{\trei_\inr}$}
\THICKLINES
\drawline\fermion[\SE\REG](\ori,\orid)[1420]
\drawline\fermion[\SW\REG](\ori,\orid)[1420]
\drawline\fermion[\NE\REG](\ori,\orid)[1420]
\drawline\fermion[\NW\REG](\ori,\orid)[1420]
\THINLINES
\global\advance\pfronty by -11000
\put(\pbackx,\pfronty){\ref{fig:scalar}}
\end{picture}
\vfil

\begin{thebibliography}{99}
\bibitem{cheng} For a review, see: H.~Y.~Cheng, \prp {\bf 158}
(1988) 1
\bibitem{thooft} H.~Georgi, H.~R.~Quinn and S.~Weinberg,
\prl {\bf 33} (1974) 451; \\
L.~Susskind, \prd {\bf 20} (1979) 2619; \\
G.~'t Hooft, in {\it Recent Developments in Gauge Theories},
ed. by G.~'t Hooft, {\it et al.} (Plenum, New York, 1980), p.135
\bibitem{kane} For a review see:\\ H.~P.~Nilles, \prp {\bf 110}
(1984) 1;\\
H.~E.~Haber and G.~L.~Kane, \prp 117 (1985) 75
\bibitem{masiero} F.~Gabbiani and A.~Masiero, \np {\bf 322}
(1989) 235 and references therein;\\
L.~I.~Bigi and F.~Gabbiani, \np {\bf 352} (1991) 309;\\
S.~Bertolini, {\it et al.}, \np {\bf 353} (1991) 591;\\
Y.~Nir and N.~Seiberg, \pl {\bf 309} (1993) 337;\\
J.~S.~Hagelin, S.~Kelley and T.~Tanaka, \np {\bf 415} (1994) 293
\bibitem{hall} L.~J.~Hall, V.~A.~Kostelecky and S.~Raby, \np
{\bf 267} (1986) 415
\bibitem{flavor} H.~Georgi, \pl {\bf 169} (1986) 231
\bibitem{leigh} M.~Dine, R.~Leigh and A.~Kagan, \prd {\bf 48} (1993)
4269
\bibitem{farhi} For a review, see: E.~Farhi and L.~Susskind, \prp
{\bf 74} (1981) 277
\bibitem{lane} S.~Dimopoulos and L.~Susskind, \np {\bf 155}
(1979) 237;\\
E.~Eichten and K.~Lane, \pl {\bf 90} (1980) 125
\bibitem{terning} T.~Appelquist and J.~Terning, \prd {\bf 50}
(1994) 2116; \\
L.~Randall, \np {\bf 403} (1993) 122
\bibitem{steve} R.~S.~Chivukula, S.~B.~Selipsky and E.~H.~Simmons,
\prl {\bf 69} (1992) 575
\bibitem{balaji} J.~Terning, \pl {\bf 344} (1995) 279; \\
B.~Balaji, Boston University preprint
BUHEP-94-14 (1994)
\bibitem{simmons} E.~H.~Simmons, \np {\bf 312} (1989) 253
\bibitem{carone} C.~D.~Carone and E.~H.~Simmons, \np {\bf 397}
(1993) 591; \\
C.~D.~Carone, E.~H.~Simmons and Y.~Su, \pl {\bf 344} (1995) 287
\bibitem{georgi} C.~D.~Carone and H.~Georgi, \prd {\bf 49} (1993) 1427
\bibitem{sekhar} R.~S.~Chivukula, A.~G.~Cohen and K.~Lane,
\np {\bf 343} (1990) 554; \\
T.~Appelquist, J.~Terning and L.~C.~R.~Wijewardhana, \prd {\bf 44}
(1991) 871
\bibitem{samuel} S.~Samuel, \np {\bf 347} (1990) 625
\bibitem{tcsusy} M.~Dine, W.~Fischler and M.~Srednicki,  \np {\bf 189}
(1981) 575; \\
S.~Dimopoulos and  S.~Raby, \np {\bf 192} (1981) 353; \\
M.~Dine and M.~Srednicki,  \np {\bf 202} (1982) 238; \\
A.~J.~Buras and T.~Yanagida,  \pl {\bf 121} (1983) 316
\bibitem{dine} M.~Dine, A.~Kagan and S.~Samuel, \pl {\bf 243} (1990)
250;\\
A.~Kagan and S.~Samuel, \pl {\bf 270} (1991) 37
\bibitem{kagan} A.~Kagan and S.~Samuel, \pl {\bf 252} (1990) 605;\\
A.~Kagan and S.~Samuel, \intl {\bf 7} (1992) 1123
\bibitem{kaplan} D.~B.~Kaplan, \np {\bf 365} (1991) 259
\bibitem{kagan1} A.~Kagan, in {\it Proceedings of the 15th Johns Hopkins
Workshop on Current Problems in Particle Theory}, ed. by G.~Domokos and
S.~Kovesi-Domokos (World Scientific, Singapore, 1992), p.217
\bibitem{kagan2} D.~Atwood, A.~Kagan and T.~G.~Rizzo,
preprint SLAC-PUB-6580 (1994), hep-ph/9407408; \\
A.~Kagan, preprint SLAC-PUB-6626 (1994), hep-ph/9409215
\bibitem{lahanas} See {\it e.g.} A.~B.~Lahanas and D.~Wyler, \pl {\bf 122}
(1983) 258;\\
A.~Masiero, D.~V.~Nanopoulos and K.~Tamvakis, \pl {\bf 126} (1983) 337;\\
W.~Buchmuller and D.~Wyler, \pl {\bf121} (1983) 321
\bibitem{peskin} M.~E.~Peskin and T.~Takeuchi, \prl {\bf 65} (1990)
964;\\
See also: \\
M.~Golden and L.~Randall, \np {\bf 361} (1991) 3; \\
B.~Holdom and J.~Terning, \pl {\bf 247} (1990) 88; \\
A.~Dobado, D.~Espriu, and M.~Herrero, \pl {\bf 255} (1991) 405; \\
M.~Peskin and T.~Takeuchi, \prd {\bf 46} (1992) 381; \\
P. Langacker, U. of Pennsylvania preprint UPR-0624-T (1994),
hep-ph/9408310
\bibitem{grisaru} L.~Girardello and M.~T.~Grisaru, \np {\bf 194} (1982)
65
\bibitem{manohar} A.~Manohar and H.~Georgi, \np {\bf 234} (1984) 189; \\
H.~Georgi and L.~Randall, \np {\bf 276} (1986) 241; \\
H.~Georgi, \pl {\bf 298} (1993) 187
\bibitem{top} CDF Collaboration (F. Abe, {\it et al.}), preprint
FERMILAB-PUB-95-022-E (1995), hep-ex/9503002; \\
D0 Collaboration (S. Abachi, {\it et al.}), preprint
FERMILAB-PUB-95-028-E (1995), hep-ex/9503003
\bibitem{data} Particle Data Group, \prd {\bf 50} (1994) 1173
\bibitem{dugan} M.~Dugan, B.~Grinstein and L.~Hall, \np {\bf 255}
(1985) 413
\bibitem{rosner} See {\it e.g.} J.~L.~Rosner, preprint EFI-94-25
(1994), hep-ph/9407257
\bibitem{ndm} J.~Ellis, S.~Ferrara and D.~V.~Nanopoulos, \pl {\bf 114}
(1983) 231; \\
J.~Polchinski and M.~Wise, \pl {\bf 125} (1983) 393; \\
W.~Buchmuller and D.~Wyler in Ref.~\cite{lahanas}
\bibitem{wolfenstein} L.~Wolfenstein, \prl {\bf 51} (1983) 1945
\bibitem{epsilon} See {\it e.g.} H.~Georgi,
{\it Weak Interactions and Modern Particle Theory} \\
(Benjamin/Cummings, Menlo Park, CA, 1984), p.158
\bibitem{spcp} A.~Pomarol, \prd {\bf 47} (1992) 273; \\
L.~Hall and S.~Weinberg, \prd {\bf 48} (1993) 979
\bibitem{sanda} I.~I.~Bigi, {\it et al.}, in {\it CP Violation}, ed. by
C.~Jarlskog (World scientific, Singapore, 1989), p.175
\bibitem{bexp} W.~Schmidt-Parzefall, in {\it CP Violation and Beauty
Factories}, ed. by D.~B.~Cline and A.~Fridman
(The New York Academy of Sciences, New York, 1991), p.281
\bibitem{buras} See {\it e.g.} A.~J.~Buras, preprint MPI-PHT-95-30
(1995), hep-ph/9504269
\bibitem{cern} NA31 Collaboration (G.~D.~Barr, {\it et al.}),
\pl {\bf 317} (1993) 233
\bibitem{fermilab} E731 Collaboration (L.~K.~Gibbons {\it et al.}),
\prl {\bf 70} (1993) 1203

\end{thebibliography}
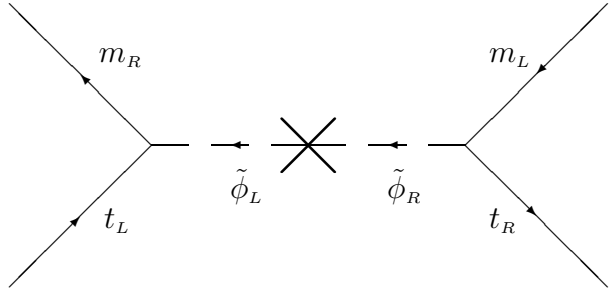
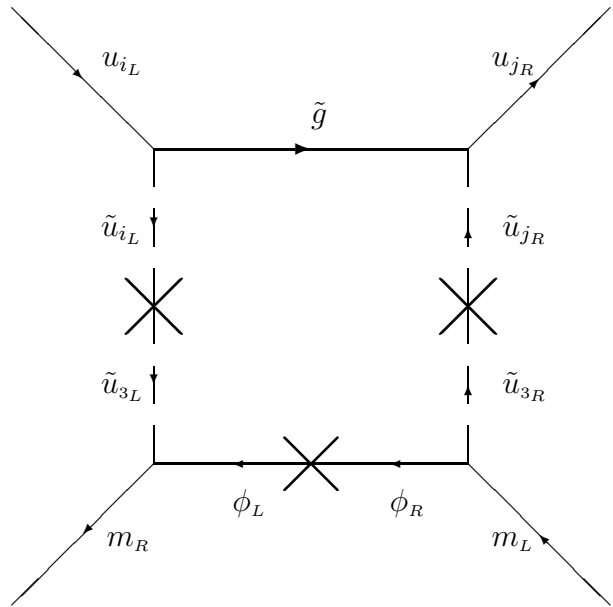
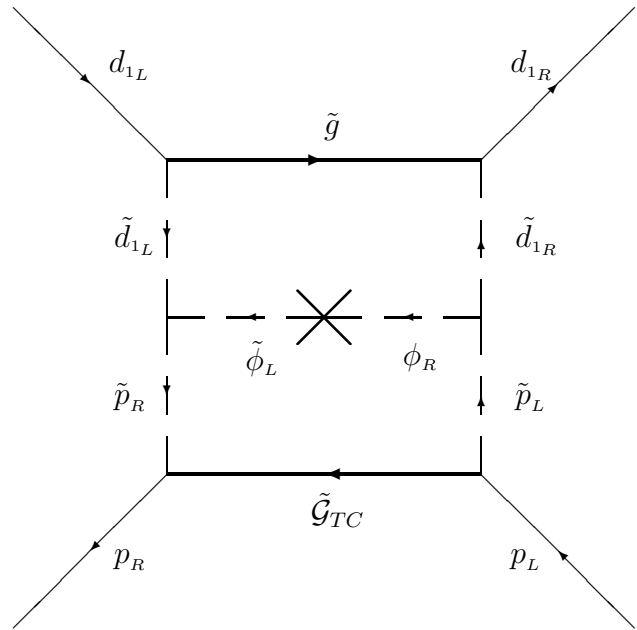
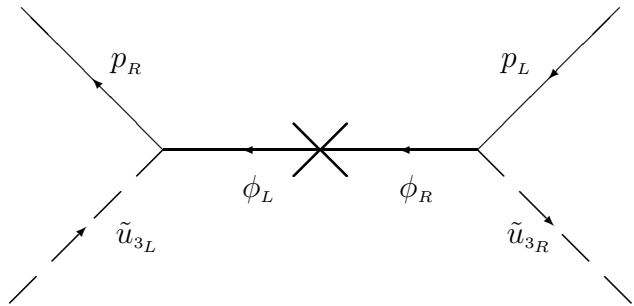
\end{document}